\documentclass[12pt]{article}
\usepackage{epsfig}

\catcode`@=11
\catcode`@=12

\newcommand{\bref}[1]{(\ref{#1})}

\newcommand{\be}{\begin{equation}}
\newcommand{\ee}{\end{equation}}
\newcommand{\bea}{\begin{eqnarray}}
\newcommand{\eea}{\end{eqnarray}}
\newcommand{\ba}{\begin{array}}
\newcommand{\ea}{\end{array}}

\def\bbox{{\,\lower0.9pt\vbox{\hrule \hbox{\vrule height 0.2 cm
\hskip 0.2 cm \vrule height 0.2 cm}\hrule}\,}}
\newcommand{\dsl}{\pa \kern-0.5em /}

\newcommand{\pa}{\partial}
\renewcommand{\t}{\theta}

\def\ka{K{\"a}hler }

\newsavebox{\DSLASH}
\sbox{\DSLASH}{$D$\hspace{-2.4mm}/}

\newsavebox{\PSLASH}
\sbox{\PSLASH}{$p$\hspace{-2.4mm}/}

\newsavebox{\PASLASH}
\sbox{\PASLASH}{$\partial$\hspace{-2.4mm}/}

\newsavebox{\LSLASH}
\sbox{\LSLASH}{$l$\hspace{-1.7mm}/}

\newsavebox{\QSLASH}
\sbox{\QSLASH}{$q$\hspace{-2.4mm}/}

\newsavebox{\KSLASH}
\sbox{\KSLASH}{$k$\hspace{-2.4mm}/}

\newsavebox{\ASLASH}
\sbox{\ASLASH}{$A$\hspace{-2.4mm}/}

\def\a{\alpha}
\def\b{\beta}
\def\g{\gamma}
\def\d{\delta}
\def\e{\epsilon}
\def\s{\sigma}

\def\part{part\'{\i}cules }

\def\otaula{\begin{tabular}}
\def\ctaula{\end{tabular}}
\def\ovec{\left( \begin{array}{c}}
\def\cvec{\end{array} \right)}
\def\omat{\left( \begin{array}{cc}}

\def\3mitj{${3 \over 2}$ }
\def\mig0{({1 \over 2},0)}
\def\0mig{(0,{1 \over 2})}

\def\undos{{1 \over 2}}
\def\espai{\;\;\;\;\;\;\;\;\;\;\;}

\def\avall{\vspace{6 mm}}

\def\pamu{\partial_{\mu}}

\def\amu{A_{\mu}}

\def\met{g_{\mu\nu}}

\def\tth{\tilde{\theta}}
\def\tphi{\tilde{\phi}}
\def\tpsi{\tilde{\psi}}

\def\L{\Lambda}

\def\l{\lambda}

\def\G{\Gamma}

\def\ie{{\it i.e. }}

\def\ut{\underline{3}}
\def\uu{\underline{1}}
\def\ud{\underline{2}}
\def\uz{\underline{0}}
\def\ur{\underline{r}}

\begin{document}


\begin{titlepage}
\vfill
\begin{flushright}
UB-ECM-PF-01/08\\

\end{flushright}
\vfill

\begin{center}
\baselineskip=16pt
{\Large\bf  
D6-branes wrapping \ka four-cycles}
\vskip 0.3cm
{\large {\sl }}
\vskip 10.mm
{\bf ~Joaquim Gomis$^1$, Toni Mateos$^2$}\\
\vskip 1cm
{\small
Departament ECM, Facultat de F\'{\i}sica\\
Universitat de Barcelona and Institut de F\'{\i}sica d'Altes
Energies,\\
Diagonal 647, E-08028 Barcelona, Spain}\\ 
\vskip.4cm
\end{center}
\vfill
\par

\begin{abstract}

We construct supergravity duals of $D6$-branes wrapping \ka four-cycles
inside a Calabi-Yau threefold, $CY_3$. We obtain the purely gravitational
M-theory description, which turns out to be a Calabi-Yau fourfold, $CY_4$.
We also analyze the dynamics of a probe $D6$ in this background.

\end{abstract}

\vfill{
 \hrule width 5.cm
\vskip 2.mm
{\small
\noindent $^1$ E-mail: gomis@ecm.ub.es \\
\noindent $^2$ E-mail: tonim@ecm.ub.es 
}}

\end{titlepage}


\section{Introduction and Results}
\indent

Recently there has been some interest to study supergravity duals of D-branes
wrapping SUSY cycles of special holonomy manifolds, 
see for example \cite{Maldacena:2001yy}-\cite{Bertolini:2001ma}. These 
studies are useful because they give information about the non-perturbative 
structure of SUSY gauge theories in various dimensions. For the case of
D6-branes, the M-theory description is purely gravitational 
and
it is interesting for a number of reasons. For example, in 
\cite{Acharya:2000gb}-\cite{Atiyah:2001qf} the M theory 
description in terms of a $G_2$ manifold 
was used to study some aspects of the IR dynamics of 4d ${\cal N}=1$ SYM

In this letter we study D6 branes wrapping 
\ka four-cycles, X, inside a Calabi-Yau threefold, $CY_3$. 
At low energies we would have an ${\cal N}=2$ twisted SUSY gauge theory
\cite{Bershadsky:1996qy} in 3 
dimensions, if we could put aside the
problems of decoupling gravity and massive string modes of D6 branes.
The purely gravitational M-theory description is given in 
terms of a Calabi-Yau four-fold, $CY_4$, consisting of
a four dimensional bundle over X.
We have constructed a one parameter family of metrics, parametrized by 
the size of the blown-up 4 cycle,
$l$,
for this space using eight-dimensional supergravity \cite{Salam:1985ft}. 
These metrics are 
asymptotically conical, and their constant radius hypersurfaces
consist of a U(1) bundle over $S^2 \times  X$. 
For $l\neq 0$ the conical singularity is resolved.
\footnote{These metrics were found in \cite{Cvetic:2000db}
 from a completely different 
approach.}
This construction exemplifies the uplift from a manifold of with SU(3)
holonomy in type IIA to a manifold with SU(4) holonomy in M theory
\cite{Gomis:2001vk}.

We reduce the metric along the Killing vector associated to a U(1) 
isometry, and we obtain a bosonic type IIA solution with a
ten-dimensional metric, a dilaton and a RR one-form.
The metric presents a curvature singularity at the place where
the $D6$ is, but it is a {\it good} singularity in the sense of
\cite{Maldacena:2001yy}.
On the other hand, the dilaton diverges at 
infinity, where classical string theory is no longer
applicable. It would be interesting to find a solution 
with a finite string coupling constant, along the same lines
as in \cite{Cvetic:2001pg} \cite{Brandhuber:2001yi}.

We have also studied the dynamics of a probe $D6$-brane in this background.
The vacuum configuration corresponds to $r=0$, 
where $r$ is radial coordinate of the cone. 
In this approximation we find that the moduli space is zero-dimensional.

\section{Twisted gauge theory}

\indent 

We consider $D6$-branes wrapping a general 
\ka four-cycle inside a Calabi-Yau three-fold $CY_3$. 
This case belongs to the well-known list 
\cite{Becker:1995kb} \cite{Becker:1996ay} of supersymmetric cycles inside
manifolds with special holonomy. In particular,
our four-cycles are calibrated by the square of
the \ka form.
The condition for a gauge theory on the brane
to be supersymmetric
actually implies that there is an identification
between the spin connection on the cycle and
the gauge connection associated to the structural group
of the normal bundle \cite{Bershadsky:1996qy}.

There is a nice way of understanding the twisting through
a group theory analysis.
A configuration with a $D6$ in flat
space would have a $SO(1,6)\times SO(3)_R$ symmetry, the
last group corresponding to the transverse directions to
the worldvolume ($R$-symmetry in the low-energy effective
field theory). The number of linearly realized 
supersymmetries is 16.
Consider now that our target space is instead $R^{1,3}\times CY_3$, 
and that we wrap the $D6$ in a \ka four-cycle inside the
$CY_3$ in such a way that its flat directions fill
an $R^{1,2}\subset R^{1,3}$. 

The worldvolume symmetry is broken to 
$SO(1,2)\times SO(4) \cong SO(1,2)\times SU(2)_1 \times
SU(2)_2$. Being a \ka four-cycle, its holonomy is only
$U(2)$, that we identify with $SU(2)_2\times U(1)_1$,
the latter being a subgroup of $SU(2)_1$.

On the other hand, the  $R$-symmetry 
will be broken to a $U(1)_R$, corresponding
to the two normal directions to the $D6$ that are inside
the $CY_3$. We summarize the way the various fields 
transform in the  original and final symmetry groups 
in a table. We indicate the $U(1)$ charges in subscripts.

\avall

\begin{center}

\otaula {c||c|c||}
 & $SO(1,6)\times SO(3)_R$
          & $SO(1,3)\times \left[SU(2)_2\times U(1)_1\right]\times U(1)_R$ 
                        \\ \hline \hline

Scalars & ({\bf 1},{\bf 3}) & ({\bf 1},{\bf 1})$_{(0,0)}\oplus$
({\bf 1},{\bf 1})$_{(0,1)}\oplus$
({\bf 1},{\bf 1})$_{(0,-1)}$ \\ \hline

Spinors & ({\bf 8},{\bf 2}) & ({\bf 2},{\bf 1})$_{(\undos,\undos)}\oplus$
({\bf 2},{\bf 1})$_{(-\undos,\undos)}\oplus$
({\bf 2},{\bf 1})$_{(\undos,-\undos)}\oplus$
({\bf 2},{\bf 1})$_{(-\undos,-\undos)}$  \\ \hline

Vectors & ({\bf 7},{\bf 1}) & ({\bf 3},{\bf 1})$_{(0,0)}\oplus$
({\bf 1},{\bf 2})$_{(\undos,0)}\oplus$
({\bf 1},{\bf 1})$_{(-\undos,0)}$   \\ \hline\hline

\ctaula

\end{center}
\avall

The twisting can now be understood as an identification of both
$U(1)$ groups, so that only those states neutral under $U(1)_D=
\left[ U(1)_1 \times U(1)_R\right]$ survive. This gives two Weyl
fermions, one scalar and one vector, which is precisely the
field content of an ${\cal N}=2$ $D=3$ SUSY theory. Later, from a supergravity
point of view, we will see that these are the spinors naturally
selected from the requirement of our solutions to be supersymmetric.

\section{BPS equations in D=8 gauged supergravity}

\indent

The aim of this section is to construct a supergravity
solution describing the 
aforementioned $D6$-brane configurations. We will work
with eight dimensional supergravity, since for $D6$-branes
one needs to give seven-dimensional boundary conditions to
the fields.
Our framework will be maximal $D=8$ gauged supergravity,
obtained in \cite{Salam:1985ft} by dimensional reduction of
$D=11$ on an $SU(2)$ manifold. We proceed to
very briefly mention their results and explain our notations.

Following the usual conventions, we will use greek characters
to describe curved indices and latin ones to describe
flat ones. Also the D=11 indices are split in $(\mu,\alpha)$
or $(a,i)$, the first ones in the D=8 space while
the second ones in the $S^3=SU(2)$.
The bosonic field content consists of the usual metric $\met$
and dilaton $\Phi$, a number of forms that we will set to zero,
an $SU(2)$ gauge potential $\amu^i$, and five scalars 
parametrizing the coset $SL(3,R)/SO(3)$ through the unimodular
matrix $L^i_{\a}$. Finally, the fermionic content consists
of a 32-components gaugino $\psi_{\mu}$ and a dilatino $\chi_i$.

We will need to make use of the susy transformations for
the fermions
\be \label{susy1}
\d\psi_{\rho}=D_{\rho}\e+{1\over 24}e^{\Phi}
F_{\mu\nu}^i\G_i\left(\G_{\rho}^{~\mu\nu}-
10 \d_\rho^{~\mu}\G^\nu\right)\e-
{g\over 288}e^{-\Phi}\e_{ijk}\G^{ijk}\G_{\rho}T\e
\ee\be \label{susy2}
\d\chi_i=\undos\left(P_{\mu ij}+{2\over 3}\d_{ij}
\pamu\Phi\right)\G^j\G^\mu\e-{1\over 4}e^{\Phi}
F_{\mu\nu i}\G^{\mu\nu}\e-{g\over 8}\left(
T_{ij}-\undos \d_{ij}T\right)\e^{jkl}\G_{kl}\e
\ee

The definitions used in this formulae are
\be 
D_{\mu}\e=\left(\pamu+{1\over 4}w_{\mu}^{ab}
\G_{ab} + {1\over 4}Q_{\mu ij}\G^{ij}\right)\e
\label{key}
\ee
\be
P_{\mu ij}+Q_{\mu ij} \equiv L_i^\a\left(\d_\a^{~\b}
\pamu - g \e_{\a\b\g} A^{\g}_\mu\right)L_{\b j}
\ee 
\be
T^{ij}=L^i_\a L^j_\b \d^{\a\b} \espai T=\d_{ij}T^{ij}
\ee
Notice that $SU(2)$ indices are raised and lowered
where $\amu^{\g}=L_i^\g\amu^i$. Finally we choose
the usual $\g$-matrices representation given by
\be
\G^a=\g^a \otimes I \espai \G^i=\g_9 \otimes \s^i
\ee
with $\g^a$ are any representation of the $D=8$ Clifford
algebra, $\g_9=i\g^0\cdots \g^7$, and $\s^i$ are
the usual $SU(2)$ Pauli matrices.

We proceed now to obtain our solutions. 
Since we look for purely bosonic SUSY backgrounds,
we must make sure that the susy transformation
of the fermions \bref{susy1}\bref{susy2} vanishes.
If we impose that the first term in \bref{susy1} vanishes, 
\ie $D_{\mu}\e=0$, we will obtain the 
twisting mentioned in the last section.
The first immediate condition that we get 
is that the metric in the four cycle must
necessarily be Einstein \cite{Gauntlett:2001ng}, so that 
\be
R_{ab}=\L g_{ab} \espai \L=cte \label{einstein}
\ee
Inspired by the
case in which the four-cycle is $CP_2$, 
we take the  metric normalized in such a way 
that \footnote{See next section for 
a discussion about the case $\L<0$.} $\L=6$. 
We then make the following 
ansatz for the $D=8$ metric
\be
ds^2_{(8)}=e^{2f(r)}dx^2_{(1,2)}+
e^{2h(r)}ds^2_{cycle}+dr^2
\ee

Now, guided by our discussion
in the last section, we complete our ansatz by 
switching on only one of the
$SU(2)_R$ gauge fields, $\amu^3$, so that we
break $R$-symmetry to $U(1)_R$, and one of the
scalars in $L^i_\a$. This matrix can therefore
be brought to \cite{Edelstein:2001pu}
\be
L^i_\a=diag(e^\l,e^\l,e^{-2\l})
\ee
Indeed, $\l$ parametrizes the Coulomb branch of the gauge theory.
We choose vielbeins for the
four-cycle such that the \ka structure takes the form
$J=e^0\wedge e^3+ e^1\wedge e^2$.
In this basis, $D_{\mu}\e=0$ further implies
the following identification
between the $R$-symmetry gauge field and the four-cycle
spin connection
\be
\amu^3=-{1\over 2g}w_{ab}J^{ab} \espai \Rightarrow \espai
F^3=dA^3=-{6\over g}\,J \label{a}
\ee
and the following projections on the
supersymmetry spinor \footnote{Every time we
write down a concrete index, we will put a 
subscript only if it is flat. So that indices in
\bref{a} are curved while those in (\ref{b},\ref{c})
are flat. Also, $\{0,1,2,3\}$ label coordinates
in the four-cycle.}
\be
\g^{\ur}\e=\e \label{b}\ee\be
\g^{\uu\ud}\e = \g^{\uz\ut} \e =\G^{\uu\ud}\label{c}\e
\ee
It is now straightforward to check that the only surviving
spinors are precisely the ones that we mentioned in the last
section. Finally,
the remaining information that  we can extract 
from our BPS equations is in the following set of coupled first-order
differential equations for the functions of our ansatz $f(r)$,
$h(r)$, for the dilaton $\Phi(r)$ and for the excited scalar $\l(r)$

\be
3f'=\Phi'={g\over 8}e^{-\Phi}(e^{-4\l}+2e^{2\l})
-{6 \over g}e^{\Phi-2h-2\l} \label{bps1}
\ee
\be
h'={g\over 24}e^{-\Phi}(e^{-4\l}+2e^{2\l})
+{4 \over g}e^{\Phi-2h-2\l}\label{bps2}
\ee\be
\l'={g\over 6}e^{-\Phi}(e^{-4\l}-2e^{2\l})
+{4 \over g}e^{\Phi-2h-2\l}\label{bps3}
\ee

\section{Solutions of the BPS equations}

\indent

For the case in which the scalar $\l$
is constant, we could obtain the following
exact solution
of the BPS equations (\ref{bps1},\ref{bps2},\ref{bps3})
\be
e^{2\Phi}={9 g^2 \over 512} \, r^2 \espai
e^{2f}=C \, r^{{2\over 3}} \espai e^{2h}={27\over 16}r^2
\espai e^{6\l}=2
\ee

There are two arbitrary integration constants. One of them is not
shown explicitely, since it just amounts to a shift in
the coordinate $r$. The other one is $C$, appearing
in the solution for $f(r)$. 

Note that if we had taken a negative value for $\Lambda$
in \bref{einstein}, the only difference would have been
a change of sign in all last terms containing ${1\over g}$.
This translates into a change of sign in the solution for
$\l$ to $e^{6\l}=-2$. Hence, there is no supersymmetric
solution for the cases $\Lambda<0$. 

One can now lift this solution
to the original $D=11$ supergravity by using the dictionary
of \cite{Salam:1985ft}. After performing a suitable redefinition
of the radial variable, we obtain

\be
ds^2_{(11)}=dx^2_{(1,2)}+2 dr^2 + 
{1\over 4}r^2(d\theta^2+sin^2\theta d\phi^2)+
{3\over 2}r^2 ds^2_{cycle} + \undos r^2 \sigma^2
\label{singular}\ee
where \footnote{These metrics were obtained in \cite{Cvetic:2000db}
in a completely different approach. Here we follow their
notation.}
\be
\sigma = d\psi -\undos cos\theta d\phi+ \tilde{A}_{[1]}
\label{sigma}\ee
Here we have defined 
$\tilde{A}_{[1]}={g\over 2} A^3_{[1]}$, so that we have 
$d\tilde{A}_{[1]}=3J$. The periodicities of the Euler
angles are $0\leq \t \leq \pi$, $0\leq \phi \leq 2\pi$,
whereas the periodicity of $\psi$ depends on which 
particular four-cycle we choose, and we leave this issue
for the particular examples.

The M-theory solution has
the topology of $R^{1,3}\times CY_4$, the Calabi-Yau
four-fold being a $C^2/Z_n$ bundle over the \ka four-cycle
(again, $n$ depends on the particular four-cycle chosen).
This is one of the lifting examples of  \cite{Gomis:2001vk}
where one goes from $SU(3)$ 
holonomy in type IIA to $SU(4)$ in M-theory.

Our metric describes a cone, with $r=cte$ hypersurfaces
described by a $U(1)$ bundle over the base 
$S^2\times X$. The particular fibration will
depend again on the four-cycle chosen. Altogether, it
forms a eight-dimensional Ricci-flat \ka metric, and
is therefore a vacuum solution of the D=11 equations. 

Note that our metric has a conical singularity at $r=0$,
where the fiber, the $S^2$ and the four-cycle collapse to
a point.
One can now try to resolve this singularity by
obtaining solutions in which at least one of the
factor spaces in the base of the cone remains finite
for $r\rightarrow 0$. This can be done here
by dropping the assumption that the scalar $\l$ is constant.
Perform the following change of variables
from the old $r$ in the BPS equations to a new one $R$
\be
{dr\over dR}=\left({gR\over 4}\right)^{\undos} U^{-{5\over 12}}(R)
\label{change}
\ee
where
\be
U(R)= {3 R^4 + 8 l^2 R^2 + 6 l^4 \over 6(R^2 +l^2)^2}
\ee
Now, a whole family of solutions parametrized by the constant
$l$ is given by
\be \label{sol1}
e^{6\l(R)}=U^{-1}(R) \espai e^{4f(R)}={g^2\over 16} R^2 U^{1\over3}(R)
\ee\be e^{2\Phi(R)}=\left({g R\over 4}\right)^3 U^{\undos}(R) \espai
e^{2h(R)}={3g\over 8} R U^{1\over6}(R) (R^2+l^2)\ee
Repeating the lifting process to M-theory, the new eleven 
dimensional metric turns out to be
\be
ds^2_{11}=dx^2_{(1,2)} + ds^2_{(8)} \label{gib1}
\ee
\be
ds^2_{(8)}=U^{-1}(R) dR^2 + {1\over 4}R^2(d\theta^2+sin^2\theta d\phi^2)+
{3\over 2}(R^2+l^2) ds^2_{cycle} + U(R) R^2 \sigma^2 \label{probeta}
\ee
Note that for $l=0$ this collapses to our first solution \bref{singular}.
On the other hand, for $l\neq 0$ the four-cycle has blown-up, and its
size remains finite at $R\rightarrow 0$, although the $S^2$ and
the $U(1)$ fiber still collapse.
Nevertheless, recall \cite{Cvetic:2001ma} that the 
the condition for local regularity in this limit implies that
{\it at most one} of the factors in the base of the $U(1)$ fiber
can collapse. Our manifold is therefore locally 
regular. Globally, it will depend on the four-cycle chosen,
as the following examples show.

\avall

{\bf Example I:} Consider the choice of
a $CP_2$ four-cycle inside a $CY_3$.
The normal directions to the $CP_2$ must form an holomorphic line
bundle, and they are completely classified by their 
first Chern class. In order to obtain a Calabi-Yau, we must
therefore take an $O(-3)$ bundle over the $CP_2$.
We provide the $CP_2$ with the standard unit
Fubini-Study metric, which is 
\be
ds^2_{CP_2}={1\over (1+\rho^2)^2}d\rho^2 + 
{\rho^2\over (1+\rho^2)^2}\sigma_3^{~2}
+ {\rho^2\over 1+\rho^2}\sigma_1^{~2} +{\rho^2\over 1+\rho^2}\sigma_2^{~2} 
\label{vielbeins} 
\ee
where $\s_i$ are the $SU(2)$ left-invariant one forms
normalized such that $d\s_i=\e_{ijk}\s_j\s_k$.
This metric is Einstein, with $R_{ab}=6\,g_{ab}$ as required.
When we plug this metric in our M-theory solution \bref{probeta},
we obtain that $\tilde{A}_{[1]}=-{3\over 2}\rho e_3$. We substitute
this in \bref{sigma} and, applying the arguments in \cite{Cvetic:2001ma}, 
we see that the maximum range of the $U(1)$ fiber angle must
be restricted to $(\Delta \psi)_{max}=\pi$ instead of the
normal $2\pi$. We have a $CP_2$ bolt at the origin.
This is why the $U(1)$ fibers over $S^2$ do not describe an
$S^3$ (viewed as a Hopf fibration), but an $S^3/Z_2$.

\avall

{\bf Example II:} We give now an example in which the
four-cycle is taken an $S^2\times S^2$. For the metric to
be Einstein both spheres need to have the same radius.
Finally, in order to normalize them such that 
$R_{ab}=6\,g_{ab}$, their radius must be $r^2=1/6$, so that
\be
ds^2_{~S^2\times S^2}={1\over 6} (d\t_1^{~2}+sin^2\t_1 d\phi_1^{~2})
+ {1\over 6} (d\t_2^{~2}+sin^2\t_2 d\phi_2^{~2})
\ee
Now $\tilde{A}_{[1]}=\undos\left[ cos\t_1 d\phi_1+cos\t_2 d\phi_2\right]$ 
so, unlike before, 
this allows $(\Delta \psi)_{max}=2\pi$. Hence, topologically,
the manifold is a $C^2$ bundle over $S^2 \times S^2$.

\section{Type IIA Analysis}

\subsection{Compactification}

\indent

In order to obtain a type $IIA$ supergravity description
of our wrapped $D6$-branes, and in order to put a probe
in this background, one can try to reduce our M-theory
solution to ten dimensions. Since the metric \bref{probeta}
has a $U(1)$ isometry, with killing vector $\partial_{\psi}$,
one can choose that direction as the M-theory circle.
In order to obtain a ten dimensional metric in the string
frame, we make the KK ansatz
\be
ds^2_{11}=e^{-2\Phi/3} ds^2_{10}+e^{4\Phi/3}(d\psi+ C_{\mu}dx^{\mu})^2
\ee
from which we obtain a bosonic type $IIA$ solution with
the following values for the metric, the dilaton and the
$RR$ one-form
\be
ds^2_{10}=e^{2\Phi/3}\left[ dx^2_{1,2}+U^{-1}dr^2+{r^2\over 4} (d\t^2+
sin^2\t d\phi^2) + {3\over 2}(r^2+l^2)ds^2_{cycle} \right] + e^{2\Phi}
(A-\undos cos\t d\phi)^2 \label{tenmetric}
\ee
\be
e^{4\Phi/3}=U(r) r^2 \espai C_{[1]}=A_{[1]}-\undos cos\t d\phi
\label{dilaton}\ee
Notice that the dilaton vanishes at $r\rightarrow 0$ and
diverges at infinity, which means that one expects a good
description with classical string theory only for small
values of $r$. Essentially, this problem comes from the
fact that our $U(1)$ fiber radius 
in the eleven-dimensional metric already diverged. Obtaining
solutions with a finite circle at infinity would probably
require an analysis beyond gauged supergravity. 
A different approach, based on imposing directly the
required symmetries in the whole D=11 supergravity, 
enabled the authors of \cite{Brandhuber:2001yi}
to construct such kind of solutions.

Our metric is clearly singular at $r\rightarrow 0$.
In order to apply the criteria for good/bad singularities
of \cite{Maldacena:2001yy}, one needs to put the metric \bref{tenmetric}
in the Einstein frame, which just amounts to multiplying by
$e^{-{\Phi\over 2}}$. It can be seen that $g_{00}$ decreases (and
it is bounded) as we approach the singularity, and so we
conclude that it is a {\it good} one, properly describing
the $IR$ behaviour of the dual theory.

\subsection{Brane probe}

\indent 

As it is already standard, we can try to learn about the
physics of our solution by putting a probe brane in 
the background of the wrapped $D6$ that we have obtained.
The natural thing is to consider the probe wrapping the
same cycle, so that one can think of it a pulling one of
the $D6$ apart from the others.
The effective action for such a probe in the case of a $CP_2$ cycle is
\be
S=-\mu_6\int_{R^{1,2}\times CP_2}d^7\xi \,\, e^{-\Phi}
\sqrt{-det[G+B_{[2]}+2\pi\a'F_{[2]}]}+ \mu_6 \int_{R^{1,2}\times CP_2}
[exp(2\pi\a'F+B)\wedge \oplus_n C_{[n]}] \label{dbi}
\ee
Here $\mu_6^{-1}= {(2\pi)}^6\a'^{7/2}$, $F_{[2]}$ is the world volume
Abelian field-strength, $B_{[2]}$ would be the NS two-form,
$C_{[n]}$ the RR $n$-forms, and all fields are understood
to be pulled-back to the seven-dimensional worldvolume.

In our solution \bref{tenmetric}\bref{dilaton} we have
$B_{[2]}=0$ and only $C_{[1]}\neq 0$. In order to 
pull back our fields we choose a static gauge, in which
we identify the worldvolume coordinates $\{\xi^i,\, i=0,...,6\}$
with the space time coordinates $\{x^0,x^1,x^2,\rho,\tth,
\tphi,\tpsi\}$, the first three parametrizing $R^{1,2}$, and
the other four the $CP_2$. We will look for the vacuum configuration
and so we will set to constant the three space time coordinates
normal to the brane $\{r,\t,\phi\}$. With these choices, our
formula \bref{dbi} becomes

\be
S=-\mu_6 \,\, Vol \left[ R^{1,2} \right] \,\, \int_{CP_2}
d\rho d\tth d\tphi d\tpsi \,\, {a^{3/2} \rho^3
(a+b\rho^2)^{1/2} sin\tth \over
8 (1+\rho^2)^3} \label{vacuum}
\ee
where $a$ and $b$ are the following functions of $r$
\be
a(r)={3\over 2}r U(r)^{\undos}(r^2+l^2) \espai
b(r)={9\over 4} r^3 U(r)^{3\over 2}
\ee
Looking at the integrand, which is always positive,
we already see that its minimum is at $r=0$ where,
indeed, $S = 0$. 

The dimension of the moduli space
can be determined by looking at the kinetic terms
arising from the DBI action when one allows for the
transverse coordinates $\{r,\t,\phi\}$ to depend
on the flat worldvolume ones $\{\xi^0,\xi^1,\xi^2\}$.
The exact expression one obtains is identical to that
in \bref{vacuum} but replacing
\be Vol \left[ R^{1,2} \right] \,\,\, \longrightarrow \,\,\,
\int d\xi_1 d\xi_2 d\xi_3 \sqrt{det\left(  \d_{ij} +  \partial_i r \partial_jr+
{1\over 4} \partial_i \t \partial_j \t +
{1\over 4}sin^2\t \partial_i \phi \partial_j \phi\right)}
\ee
Here $\{\partial_i = {\partial \over \partial \xi^i},\, i=0,1,2\}$.
Clearly, evaluating this at the minimum $r=0$ still
makes the whole expression vanish. Hence, 
In this approximation we find that the moduli space is zero-dimensional.

\vskip 6mm

{\it{\bf Acknowledgments}}

We would like to thank Jerome Gauntlett, Jaume Gomis, David Mateos,
Alfonso Ramallo and Joan Sim\'on for
useful discussions. 
This work is partially supported by
AEN98-0431 and GC 2000SGR-00026 (CIRIT). 
T.M. is supported by a fellowship from the Commissionat per 
a la Recerca de la Generalitat de Catalunya.

\end{document}